\documentclass[aps,a4paper,twocolumn,floatfix,showpacs,showkeys,groupedaddress]{revtex4}

\usepackage{bm,amsmath,amssymb,amsfonts}

\usepackage[dvips]{graphicx}

\usepackage{color}
\definecolor{vio}{rgb}{0.46,0.14,0.75}
\begin{document}

\title{\textcolor{vio}
{Absolutely continuous energy bands and extended electronic states 
in \\an aperiodic comb-shaped nanostructure}}

\author{Biplab Pal}
\email{biplabpal@klyuniv.ac.in}

\affiliation{Department of Physics, University of Kalyani, Kalyani,
West Bengal-741235, India}

\begin{abstract}
The nature of electronic eigenstates and quantum transport in a comb-shaped 
Fibonacci nanostructure model is investigated within a tight-binding framework. 
Periodic linear chains are side-attached to a Fibonacci chain, giving it the shape 
of an aperiodic comb. The effect of the side-attachments on the usual Cantor set 
energy spectrum of a Fibonacci chain is analyzed using the Green’s function 
technique. A special correlation between the coupling of the side-attached chain 
with the Fibonacci chain and the inter-atomic coupling of the Fibonacci chain 
results in a dramatic triggering of the fragmented Cantor set energy spectrum into 
multiple sets of continuous sub-bands of extended eigenstates. The result is valid 
even for a disordered comb and turns out to be a rare exception of the 
conventional Anderson localization problem. The electronic transport thus can be 
made selectively ballistic within desired energy regimes. The number and the width 
of such continuous sub-bands can be easily controlled by tuning the number of 
atomic sites in the side-coupled periodic linear chains. This gives us a scope of 
proposing such aperiodic nanostructures as potential candidates for prospective 
energy selective nanoscale filtering devices.
\end{abstract}

\maketitle
%%%%%%%%%%%%%%%%%%%%%%%%%%%%%%%%%%%%%%%%%%%%%%%%%%%%%%%%%%%%%%%%%%%%%%%%%%%%%%%%%
\section{Introduction}
\label{sec1}
Fano-Anderson effect is an interesting phenomenon exhibited by excitonic 
states in a tight binding lattice when a discrete, bound state `interacts' 
with a continuum~\cite{mirosh,flach,orellana1,franco,aligia1,kobayashi,
orellana2,pouthier,zhang1,arunava1,arunava2,fano}. The effect is explicitly 
seen when one considers transmission spectrum of one dimensional quantum 
wire (QW) systems with a single quantum dot (QD), or a cluster of them 
is attached to the QW from one side~\cite{orellana2,pouthier,arunava1,arunava2}. 
Needless to say, the studies so far has gone well beyond mere theoretical 
interest, thanks to the present advanced stage of lithographic techniques and 
nanotechnology, and has incorporated the studies of nonequilibrium dynamics 
in optical transition~\cite{haupt}, quantum simulation~\cite{hansen} and 
tunneling in a Kondo hole system~\cite{zhu} to name a few. 

Inspite of the considerable volume of work existing in this field, a 
practically unaddressed issue, to the best of our knowledge, is how 
seriously does the presence of {\it bound states} caused by the attachment 
(from one side) of a QD or an assembly of them, influence a singular 
continuous spectrum. A singular continuous spectrum is the hallmark of 
what are known as quasiperiodic systems~\cite{kohmoto1,sokoloff,kohmoto2,
macia,naumis} such as the Fibonacci class of substitutionally generated atomic 
chains or, the array of Harper potentials (Aubry-Andre chains). Due to the 
fragmented, Cantor like character of the energy spectra offered by these 
systems, the bound states can, in principle, occupy any local gap inside 
the spectrum and may de-stabilize the fundamental character of the entire 
energy spectrum. A central motivation of the present work is to look into 
such an issue. 

We intend to work with a golden mean Fibonacci chain where finite arrays 
of QDs are side-grafted to every atomic site, giving the lattice an 
aperiodic comb structure. The model has particular relevance in the 
context of recent studies on spectral properties of excitons in nanoparticle 
array~\cite{dalnegro1,dalnegro2,dalnegro3,dalnegro4,kaputkina}. The 
comb-structure enables us to examine the interference of Fano lineshapes 
and the effect of the side coupled dots on the energy spectrum and transport 
properties of such a system. The presence of multiple, closely spaced gaps 
and a multifractal spectrum, when coupled to discrete, bound states from 
one side, offers a possibility of controlling the gaps, and may be even 
the basic character of the electronic states. In the present day of 
nanotechnology and revolution in fabricating tailor made lattices this can 
even open up the possibility of designing novel QD devices. These aspects 
prompt us to undertake the present study.

We find extremely interesting results. With a golden mean Fibonacci chain 
having side coupled QD arrays, the presence of an infinite number of grafted 
chains enables one to control the transmission properties at all scales of 
energy by suitably changing the number of atoms in the side coupled array. 
The most important result is that, under certain special conditions the 
entire energy spectrum of the infinite system turns out to have absolutely 
continuous subbands of energy. The occurrence of these continuous subbands 
is triggered by a certain special relationship between the off-diagonal 
elements of the Hamiltonian. Surprisingly, this is something that is not 
expected in any conventional localization scheme, such as the Anderson 
localization~\cite{anderson}, where the single particle states get localized 
in a disordered array of potentials irrespective of the numerical values 
of the parameters of the Hamiltonian. Our results therefore present a 
strong case of a violation of this {\it universality} of Anderson localization.   

The system to be described by a tight-binding Hamiltonian, is depicted in 
Fig.~\ref{system}. We have used real space renormalization group (RSRG) 
decimation scheme~\cite{arunava3} and Green's function technique to analyze 
the energy spectrum and the nature of eigenstates of the system. 
The side-attached QD arrays play a crucial role in generating and 
controlling the number of continuous energy bands. Extended character 
of the electron states populated in the bands are reflected in a prominent 
enhancement of two-terminal quantum transport through the system.

In what follows, we describe the results obtained by us. In section~\ref{sec2} 
we present our model and the mathematical methods used for handling the 
problem. The effect of the side-attachment on the energy spectrum of the 
system and how to obtain the condition for generating absolutely continuous 
bands of extended states is discussed in section~\ref{sec3}. Section~\ref{sec4} 
presents the two-terminal transmission characteristics of the system, 
and finally in section~\ref{sec5} we draw our conclusion with possible 
applications for future study.  
%%%%%%%%%%%%%%%%%%%%%%%%%%%%%%%%%%%%%%%%%%%%%%%%%%%%%%%%%%%%%%%%%%%%%%%%%%%%%%%%%                    
\section{Description of the model and the method}
\label{sec2}
%******************************************************
\subsection{The system and the Hamiltonian}
\label{subsec2.1}
\begin{figure}[ht]
\centering
\includegraphics[clip,width=8cm,angle=0]{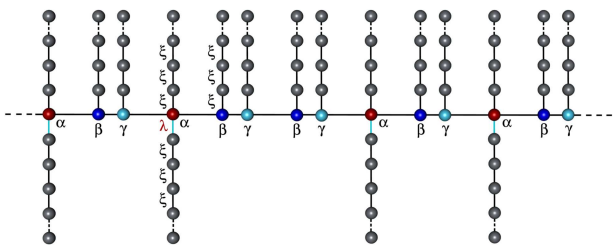}
\caption{(Color online) Schematic diagram of a model quasiperiodic 
nanostructure. Sites in the backbone are named and colored differently 
to distinguish between nearest neighbor bond environments. $\alpha$, 
$\beta$, and $\gamma$ sites are side-attached with finite size periodic 
QD arrays, while each $\alpha$ site has an extra dangling QD chain coupled 
to it through a coupling strength $\lambda$.}
\label{system}
\end{figure}

Fig.~\ref{system} illustrates our system of interest. 
We have considered a model quasiperiodic nanostructure, in which 
atomic sites are arranged following a Fibonacci sequence on a linear 
chain (the backbone) and periodic, finite sized QD arrays are side-attached 
to the backbone forming an aperiodic comb-shaped system.  
%@@@@@@@@@@@@@@@@@@@@@@@@@@@@@@@@@@@@@@@@@@@@@@@@@@@ 
%@@@@@@@@@@@@@@@@@@@@@@@@@@@@@@@@@@@@@@@@@@@@@@@@@@@
The Fibonacci chain is grown recursively by repeated application of the 
inflation rule~\cite{kohmoto1,kohmoto2} $L \rightarrow LS$ and $S \rightarrow L$, 
where $L$ and $S$ stand for two ``bonds", viz., long bond and short bond 
respectively. The first few generations are,
\begin{center}
$G_{1}$:$L$, $G_{2}$:$LS$, $G_{3}$:$LSL$, $G_{4}$:$LSLLS$
\end{center}
and so on. The number of $L$ bonds and the number of $S$ bonds bear the golden 
ratio $\tau=(\sqrt{5}+1)/2$ in the thermodynamic limit. 

Depending on the nearest neighbor bond environment, we identify three 
kinds of sites, viz., $\alpha$ (red dots), $\beta$ (dark blue dots), 
and $\gamma$ (light blue dots) -- flanked by $L$-$L$, $L$-$S$ and 
$S$-$L$ bonds respectively as shown in Fig.~\ref{system}. Periodic 
finite QD arrays of identical size (having $N+1$ QDs in each array) are 
side-attached to every $\alpha$, $\beta$, and $\gamma$ sites, 
while an extra QD array of size $N+2$ is hung from each $\alpha$ site, 
which couples to it through a coupling strength $\lambda$ as shown in 
Fig.~\ref{system}, $N$ being any positive integer. We adopt a 
tight-binding formalism and incorporate only nearest 
neighbor hopping. Within the framework of non-interacting electron 
picture, the tight-binding Hamiltonian of the system can be expressed as,
\begin{eqnarray}
{\bm{H}}={\bm{H}}_{backbone}+{\bm{H}}_{upperchains}+{\bm{H}}_{lowerchains}\nonumber\\
\qquad +{\bm{H}}_{backbone\text{-}upperchains}+{\bm{H}}_{backbone\text{-}lowerchains}
\label{hamilton}
\end{eqnarray}
where,
\begin{align*}
&{\bm{H}}_{backbone}=\sum\limits_{i}\epsilon_{i}c_{i}^{\dag}c_{i} + 
\sum\limits_{\langle ij \rangle}\big(t_{ij}c_{i}^{\dag}c_{j}+h.c.\big) \\
&{\bm{H}}_{upperchains}=\sum\limits_{\alpha,\beta,\gamma\ \text{sites}}
\bigg[\sum\limits_{\mu=1}^{N+1}\epsilon_{\mu}d_{\mu}^{\dag}d_{\mu} + 
\sum\limits_{\langle \mu\nu \rangle}
\big(t_{\mu\nu}d_{\mu}^{\dag}d_{\nu}+h.c.\big)\bigg] \\
&{\bm{H}}_{lowerchains}=\sum\limits_{\alpha\ \text{sites}}
\bigg[\sum\limits_{\mu=1}^{N+2}\epsilon_{\mu}d_{\mu}^{\dag}d_{\mu} + 
\sum\limits_{\langle \mu\nu \rangle}
\big(t_{\mu\nu}d_{\mu}^{\dag}d_{\nu}+h.c.\big)\bigg] \\
&{\bm{H}}_{backbone\text{-}upperchains}=\sum\limits_{\alpha,\beta,\gamma\ \text{sites}}
\xi\big(c_{i}^{\dag}d_{1}+d_{1}^{\dag}c_{i}\big) \\
&{\bm{H}}_{backbone\text{-}lowerchains}=\sum\limits_{\alpha\ \text{sites}}
\lambda\big(c_{i}^{\dag}d_{1}+d_{1}^{\dag}c_{i}\big)
\end{align*}
In the above, $c_{i}^{\dag}(c_{i})$ and $d_{\mu}^{\dag}(d_{\mu})$ represent the 
creation (annihilation) operators for the backbone and dangling QD arrays 
respectively. $\epsilon_{i}$ is the on-site potential of an $i$-th atomic 
site along the Fibonacci backbone of the lattice, which may take three different 
values, viz., $\epsilon_{\alpha}$, $\epsilon_{\beta}$, and $\epsilon_{\gamma}$ 
corresponding to three different kinds of sites as mentioned above. The 
nearest neighbor hopping integral $t_{ij}$ takes two values $t_{L}$ and $t_{S}$ 
for electron hopping along the $L$ and $S$ bond respectively. $\epsilon_{\mu}$ is 
the value of on-site energy at an atomic site in the dangling chains, which will 
assume a constant value equal to $\epsilon_{0}$ for all the QDs in an array. 
The hopping integral along a dangling QD array also assumes a constant value 
$t_{\mu\nu}=\xi$, and $\lambda$ is the coupling strength of the extra QD 
arrays coupled to each $\alpha$-site.
%******************************************************
\subsection{Renormalization into an effective system}
\label{subsec2.2}
The dangling chains are ``folded back" into the backbone using a RSRG decimation 
scheme~\cite{jana}. As a result, the on-site energy of 
the renormalized atomic sites gets dressed up, which we may call 
a kind of ``self energy" of the effective atomic sites. 
%@@@@@@@@@@@@@@@@@@@@@@@@@@@@@@@@@@@@@@@@@@@@@@@@@@@ 
\begin{figure}[ht]
\centering
\includegraphics[clip,width=8cm,angle=0]{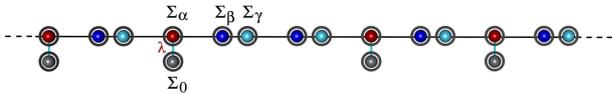}
\caption{(Color online) Schematic view of renormalized version of the 
original system. The side-attached dangling QD arrays are folded into 
effective atomic sites shown by encircled dots.}
\label{decimated system}
\end{figure}
%@@@@@@@@@@@@@@@@@@@@@@@@@@@@@@@@@@@@@@@@@@@@@@@@@@@
This is demonstrated 
pictorially in Fig.~\ref{decimated system} and the renormalized on-site 
potentials of the effective sites are given by,  
\begin{equation}
\begin{aligned}
\Sigma_{\alpha}=\epsilon_{\alpha}+\Delta\\
\Sigma_{\beta}=\epsilon_{\beta}+\Delta\\
\Sigma_{\gamma}=\epsilon_{\gamma}+\Delta\\
\Sigma_{0}=\epsilon_{0}+\Delta
\end{aligned}
\label{selfeng}
\end{equation}
In the above equations, $\Delta$ is the term which contains the entire 
information about the folded periodic finite size QD arrays and is given by,
\begin{equation}
\Delta=\dfrac{\xi (E-\epsilon_{0})U_{N-1}(x)-\xi^{2} U_{N-2}(x)}
{(E-\epsilon_{0})U_{N}(x)-\xi U_{N-1}(x)}, \quad \text{for}\ N\geq 1
\label{delta}
\end{equation}
where, $U_{N}(x)=2xU_{N-1}(x)-U_{N-2}(x)$, for $N \geq 1$ is the Chebyshev polynomial 
of second kind, with $U_{-1}=0$, $U_{0}=1$, and $x=1/2[\text{Tr}({\bm M})]$, 
${\bm M}$ being the $2 \times 2$ transfer matrix of the form,
\begin{equation*}
{\bm M}=
\left( {\def\arraystretch{1.5}\begin{array}{cc}
(E-\epsilon_{0})/\xi & -1 \\
1 & 0 \\
\end{array} } \right)
\end{equation*}
for the atomic sites in the QD arrays. 

\begin{figure}[ht]
\centering
\includegraphics[clip,width=8cm,angle=0]{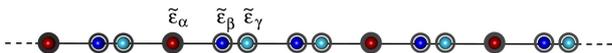}
\caption{(Color online) Effective 1-d Fibonacci array of atomic 
sites obtained by decimating the dangling renormalized QDs at the 
$\alpha$-sites in figure~\ref{decimated system}.}
\label{effective system}
\end{figure}

Now we have a renormalized system 
with effective atomic sites arranged in a Fibonacci order along the 
backbone and each effective $\alpha$-sites is now coupled to an 
effective dangling QD through a coupling strength $\lambda$ 
(Fig.~\ref{decimated system}). We can easily decimate out the 
dangling ``effective" QD, which will make the system into an 
effective 1-d Fibonacci array of renormalized atomic sites with 
renormalized on-site potentials, viz., 
${\tilde{\epsilon}}_{\alpha}=\Sigma_{\alpha}+\lambda^{2}/(E-\Sigma_{0})$, 
${\tilde{\epsilon}}_{\beta}=\Sigma_{\beta}$, and 
${\tilde{\epsilon}}_{\gamma}=\Sigma_{\gamma}$ as depicted in 
Fig.~\ref{effective system}. 
%@@@@@@@@@@@@@@@@@@@@@@@@@@@@@@@@@@@@@@@@@@@@@@@@@@@ 
%@@@@@@@@@@@@@@@@@@@@@@@@@@@@@@@@@@@@@@@@@@@@@@@@@@@
Actually the on-site potentials at 
the $\beta$ and $\gamma$ sites do not get modified, but the to make 
the notations consistent we have renamed $\Sigma_{\beta}$ and 
$\Sigma_{\gamma}$ as ${\tilde{\epsilon}}_{\beta}$ 
and ${\tilde{\epsilon}}_{\gamma}$ respectively. 

Using the Fibonacci growth-rule in the reverse direction it is simple to 
rescale any generation of the effective 1-d Fibonacci system 
into its earlier generation~\cite{arunava3}. The renormalized values of 
the on-site energy and hopping integrals 
can be expressed by the following relations:
\begin{equation} 
\begin{aligned}
&{\tilde{\epsilon}}_{\alpha}(n+1)={\tilde{\epsilon}}_{\gamma}(n)+
\dfrac{t_{L}^{2}(n)+t_{S}^{2}(n)}{E-{\tilde{\epsilon}}_{\beta}(n)}\\
&{\tilde{\epsilon}}_{\beta}(n+1)={\tilde{\epsilon}}_{\gamma}(n)+
\dfrac{t_{S}^{2}(n)}{E-{\tilde{\epsilon}}_{\beta}(n)}\\
&{\tilde{\epsilon}}_{\gamma}(n+1)={\tilde{\epsilon}}_{\alpha}(n)+
\dfrac{t_{L}^{2}(n)}{E-{\tilde{\epsilon}}_{\beta}(n)}\\
&t_{L}(n+1)=\dfrac{t_{L}(n)t_{S}(n)}{E-{\tilde{\epsilon}}_{\beta}(n)}\\
&t_{S}(n+1)=t_{L}(n)
\end{aligned}
\label{recursion1}
\end{equation}
where, $n$ indicates the stage of renormalization. 
These recursion relations are used to obtain the energy spectrums of the 
system and are presented in the next the section.
%******************************************************         
%%%%%%%%%%%%%%%%%%%%%%%%%%%%%%%%%%%%%%%%%%%%%%%%%%%%%%%%%%%%%%%%%%%%%%%%%%%%%%%%%
\section{Energy spectrums of the system}
\label{sec3}
%******************************************************
\subsection{Condition for obtaining the absolutely continuous bands}
\label{subsec3.1}
The renormalized effective 1-d Fibonacci chain is described by a 
set of three transfer matrices~\cite{kohmoto2}, 
viz., ${\bm{M_{\alpha}}}$, ${\bm{M_{\beta}}}$, 
and ${\bm{M_{\gamma}}}$. The expressions for the three transfer matrices are,
\begin{equation}
\begin{aligned}
&{\bm{M_{\alpha}}}=
\left( {\def\arraystretch{1.5}\begin{array}{cc}
(E-{\tilde{\epsilon}}_{\alpha})/t_{L} & -1 \\
1 & 0 \\
\end{array} } \right)\\
&{\bm{M_{\beta}}}=
\left( {\def\arraystretch{1.5}\begin{array}{cc}
(E-{\tilde{\epsilon}}_{\beta})/t_{S} & -t_{L}/t_{S} \\
1 & 0 \\
\end{array} } \right)\\
&{\bm{M_{\gamma}}}=
\left( {\def\arraystretch{1.5}\begin{array}{cc}
(E-{\tilde{\epsilon}}_{\gamma})/t_{L} & -t_{S}/t_{L} \\
1 & 0 \\
\end{array} } \right)
\end{aligned}
\label{matrices}
\end{equation}
and, ${\bm{M_{\alpha}}}$ and ${\bm{M_{\gamma\beta}}}=
{\bm{M_{\gamma}}}.{\bm{M_{\beta}}}$ follow an arrangement in the 
Fibonacci sequence. It can be easily verified that, if we choose 
$\epsilon_{\alpha}=\epsilon_{\beta}=\epsilon_{\gamma}=\epsilon_{0}$ and 
$t_{L} \neq t_{S}$, then the commutator 
$\left[{\bm{M_{\alpha}}},{\bm{M_{\gamma\beta}}}\right]$ vanishes 
{\it irrespective of the energy $E$}, if we set 
$\lambda=\sqrt{t_{S}^{2}-t_{L}^{2}}$. So, under this condition, when the 
transfer matrices corresponding to $\alpha$ clusters and 
the $\beta\text{-}\gamma$ clusters commute, they can be arranged 
in any desired fashion, say, for example, in a perfectly periodic pattern. 
The spectrum offered by a periodic arrangement of building blocks is an 
absolutely continuous one. Once the commutator 
$\left[{\bm{M_{\alpha}}},{\bm{M_{\gamma\beta}}}\right]$ is made to vanish, the 
entire Fibonacci lattice can be thought to be equivalent with two periodic infinite 
sub-lattices, one comprising the ``renormalized" $\alpha$-sites alone, and the 
other formed by a periodically repeating cluster of $\beta\gamma$ pairs. Each 
subsystem has its own continuous energy spectrum with a 
band-multiplicity that arises out of the side-coupled QD array. 
%@@@@@@@@@@@@@@@@@@@@@@@@@@@@@@@@@@@@@@@@@@@@@@@@@@@ 
\begin{figure*}[ht]
\centering
\includegraphics[clip,width=15cm,angle=0]{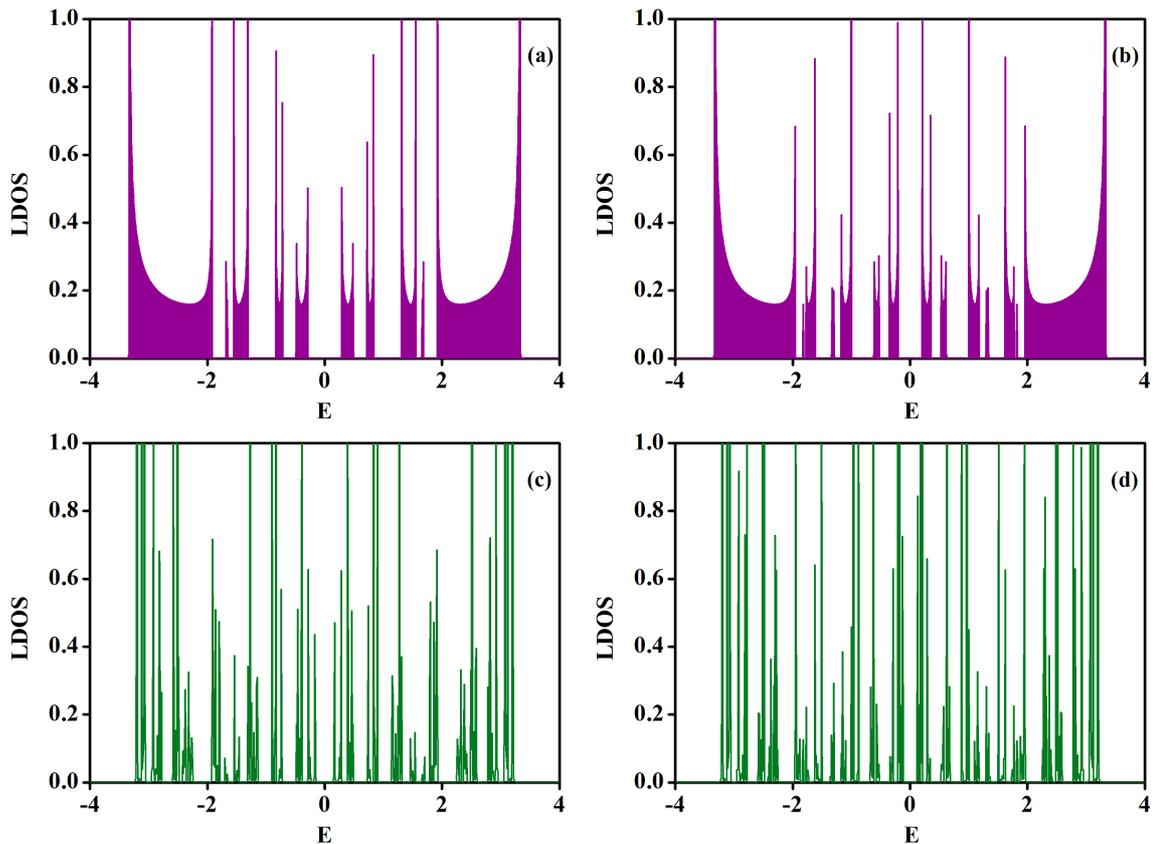}
\caption{(Color online) LDOS-$E$ spectrum at an $\alpha$-site of the system. 
The upper panel shows the LDOS-$E$ spectrum under the suitable resonance 
condition, while the lower panel represents the LDOS-$E$ spectrum for a $50\%$ 
deviation from the desired resonance condition. In (a) and (c) number of 
atomic sites in each lower chain is $5$ and in each upper chain is $4$, while 
for (b) and (d) those are set equal to $7$ and $6$ respectively. The values of 
other parameters are $\epsilon_{\alpha}=\epsilon_{\beta}=\epsilon_{\gamma}=\epsilon_{0}=0$, 
$t_{L}=1$, $t_{S}=2$, and $\xi=1$. For upper panel value of $\lambda=\sqrt{3}$, and for 
lower panel $\lambda=\sqrt{3}/2$ which is a $50\%$ deviation from the resonance value.}
\label{ldos}
\end{figure*}
%@@@@@@@@@@@@@@@@@@@@@@@@@@@@@@@@@@@@@@@@@@@@@@@@@@@
Most interestingly, 
the separate band structure arising in these two cases map onto each other, 
giving rise to a unique, absolutely continuous energy spectrum as verified 
by an extensive numerical work. This conclusively leads to the existence of a dense, 
gapless set of extended eigenfunctions when the resonance condition is satisfied. The 
fragmented, multifractal character of a typical Fibonacci quasicrystal disappears and 
all the eigenstates populating the continuous subbands are of extended character.   
%******************************************************
\subsection{Local density of states}
\label{subsec3.2}
The local density of states (LDOS) at an $\alpha$, $\beta$ or $\gamma$ is obtained by 
calculating the respective local Green's function 
$G_{00}^{i}=(E+i\delta-\epsilon_{i}^{\ast})^{-1}$, where, $i=\alpha$, $\beta$ or $\gamma$. 
$\delta$ is a very small imaginary part added to the energy $E$, and $\epsilon_{i}^{\ast}$ 
represents the fixed point value of the respective on-site potential 
as the hopping integrals flows to zero under RSRG iterations of Eq.~\eqref{recursion1}. The 
LDOS is given by, 
\begin{equation}
\rho_{i}(E)= \lim_{\delta \rightarrow 0}\,\left[-\dfrac{1}{\pi}\,
\text{Im}\,\big(G_{00}^{i}\big)\right]
\end{equation}
The results are presented in Fig.~\ref{ldos}. We have verified that the LDOS at an 
$\alpha$, or $\beta$, or $\gamma$ site turns out to be the same. So to save space we 
have presented the LDOS at an $\alpha$-site only. 
In the upper panel of Fig.~\ref{ldos}, we have shown the LDOS at an $\alpha$-site under 
the resonance condition. We have set 
$\epsilon_{\alpha}=\epsilon_{\beta}=\epsilon_{\gamma}=\epsilon_{0}=0$, 
$t_{L}=1$, $t_{S}=2$, $\xi=1$, and $\lambda=\sqrt{3}$. As we set $\lambda=\sqrt{3}$, which 
satisfies the desired condition for matrix commutation as described in Sec.~\ref{subsec3.1}, 
it triggers absolutely continuous subbands in the energy spectra (Fig.~\ref{ldos}(a) and (b)). 

We have minutely examined each of these continuum sub-bands for a very fine 
scanning of the energy interval, and the continua still persist. This confirms 
the robustness of these continuum subbands under the resonance condition. The 
values of the hopping integrals $t_{L}$ and $t_{S}$ remains non-zero over 
arbitrarily large number of RSRG iteration steps for any energy inside each of these 
continuous subbands. This proves the extended nature of the electronic 
eigenstates in each of these subbands. This fact is corroborated by an prominent 
enhancement of two-terminal transport through the system under the resonance 
condition which will be described in section~\ref{sec4}. The number and the position 
of such subbands are sensitive to number of atomic sites in the side-attached chains. 
This fact is elucidated in Fig.~\ref{ldos}(a) and (b). In Fig.~\ref{ldos}(a), 
number of sites in each lower chain is set equal to $5$ and in each upper chain is 
set equal to $4$ respectively, and for Fig.~\ref{ldos}(b) these are $7$ 
and $6$ respectively. Clearly, the number and position of subbands in the two 
cases change. 

As we turn away from the resonance condition, the continua in the energy spectra 
get destroyed, and the typical Cantor set singular continuous energy spectra of a 
quasiperiodic system returns. This is shown in the lower panel panel of 
Fig.~\ref{ldos}, where we have set $\lambda=\sqrt{3}/2$ which is a $50\%$ 
deviation from its resonance value. All other parameters remain same as in 
upper panel of Fig.~\ref{ldos}. However, isolated extended states still 
exists in such cases. These can be evaluated exactly.
%******************************************************
\subsection{The energy eigenvalue distribution}
\label{subsec3.3}
\begin{figure}[ht]
\centering
\includegraphics[clip,width=8cm,angle=0]{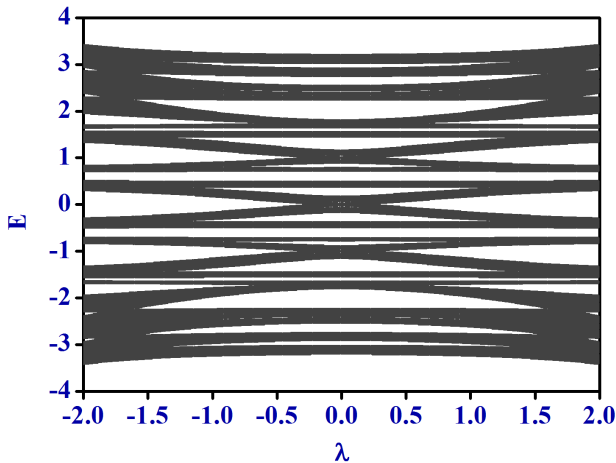}
\caption{(Color online) Distribution of energy eigenvalues spectrum of the 
proposed system as a function of the coupling 
strength $\lambda$. We have chosen 
$\epsilon_{\alpha}=\epsilon_{\beta}=\epsilon_{\gamma}=\epsilon_{0}=0$, 
$t_{L}=1$, $t_{S}=2$, and $\xi=1$. Number of sites in each side-coupled lower 
chain is set to $5$ and in each upper chain is set to $4$.}
\label{engspectra}
\end{figure}

We have explored the energy eigenvalue distribution (Fig.~\ref{engspectra}) 
as a function of the coupling strength $\lambda$ of the dangling chains with 
the $\alpha$-sites. 
%@@@@@@@@@@@@@@@@@@@@@@@@@@@@@@@@@@@@@@@@@@@@@@@@@@@ 
%@@@@@@@@@@@@@@@@@@@@@@@@@@@@@@@@@@@@@@@@@@@@@@@@@@@ 
Fig.~\ref{engspectra} is representative of an infinite system. In Fig.~\ref{engspectra} we 
observe formation of multiple bands and gaps in the energy eigenvalue spectrum as we vary the 
value of $\lambda$. For values of $\lambda$ away from $\sqrt{3}$, the spectrum shows the 
fragmentation with relatively dense packing of energy eigenvalue around certain energy 
intervals which basically are due to isolated extended states (or, at least states with a 
large localization length) clustered together. Band-crossing is, in general avoided in 
such cases. With $\lambda$ close to the resonance value of $\sqrt{3}$, absolutely 
continuous sub-bands start dominating the spectrum, and multifractality disappears.
%******************************************************
%%%%%%%%%%%%%%%%%%%%%%%%%%%%%%%%%%%%%%%%%%%%%%%%%%%%%%%%%%%%%%%%%%%%%%%%%%%%%%%%%
\section{Two terminal transmission characteristics}
\label{sec4}
\begin{figure}[ht]
\centering
\includegraphics[clip,width=8.5cm,angle=0]{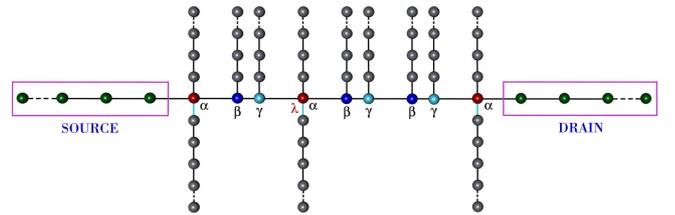}
\caption{(Color online) Schematic view of the system clamped between two semi-infinite 
ordered leads, viz., source and drain.}
\label{esebridge}
\end{figure}

To obtain the the two-terminal transmission characteristics of the system, we clamp the system 
between two semi-infinite ordered leads, viz., source and drain (Fig.~\ref{esebridge}). The 
leads are characterized by uniform on-site potential $\epsilon_{\mathcal{L}}$ and nearest 
neighbor hopping integral $\tau_{\mathcal{L}}$. 
%@@@@@@@@@@@@@@@@@@@@@@@@@@@@@@@@@@@@@@@@@@@@@@@@@@@ 
%@@@@@@@@@@@@@@@@@@@@@@@@@@@@@@@@@@@@@@@@@@@@@@@@@@@
We renormalize the system clamped between the two leads into an effective dimer 
by decimating out the internal sites selectively. To get unique recursion 
relations at the two boundary sites of the system, we consider system with odd 
generation index only. This should be appreciated that, as we interested to 
look at the behavior of transmission characteristics of a large finite size system, 
by choosing system with odd generation index only we do not lose any physics 
whatsoever. Instead of deflation rule used in Sec.~\ref{subsec2.2}, we now 
use an alternative deflation rule 
$LSL \rightarrow L^{\prime}$ and $LS \rightarrow S^{\prime}$~\cite{arunava1}. 
By using this deflation rule, staring with an $(2n+1)$-th generation Fibonacci system, 
decimation of $n$ steps converts it into an effective dimer with two renormalized 
atomic sites having effective on-site potentials 
${\tilde{\epsilon}}_{l}^{\ast}$ and ${\tilde{\epsilon}}_{r}^{\ast}$ 
(for the ``left" and ``right" edge sites) and are connected to each other by an effective 
hopping integral $t_{L}^{\ast}$. The recursion relations arises out of this 
decimation procedure are given by,
%@@@@@@@@@@@@@@@@@@@@@@@@@@@@@@@@@@@@@@@@@@@@@@@@@@@ 
\begin{figure*}[ht]
\centering
\includegraphics[clip,width=15cm,angle=0]{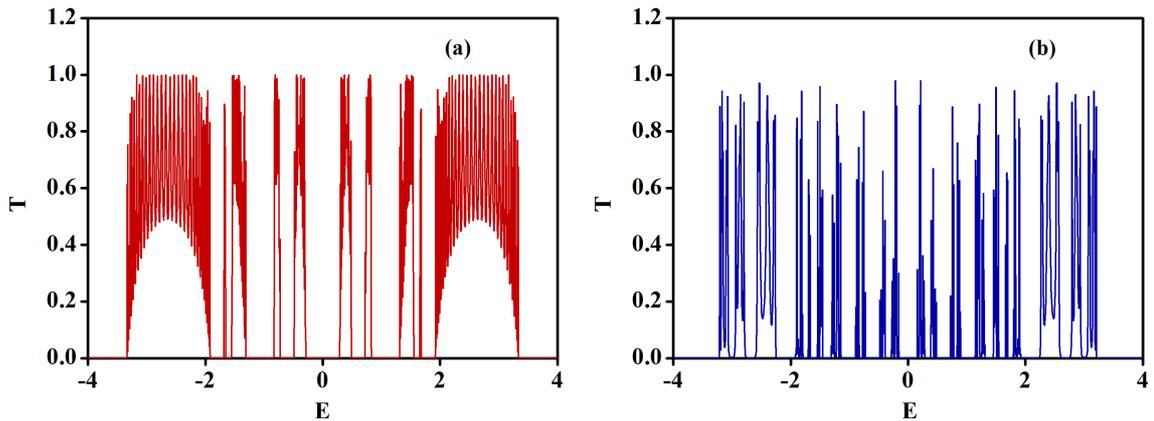}
\caption{(Color online) Transmission characteristics of a $9$-th generation system 
as a function of the energy $E$. (a) represents the transmission characteristics 
under the resonance condition ($\lambda=\sqrt{3}$) and (b) is for the $50\%$ deviation 
from the resonance condition. Number of sites in each side-coupled lowerchain is 
$5$ and in each upper chain is $4$. The values of other parameters are 
$\epsilon_{\alpha}=\epsilon_{\beta}=\epsilon_{\gamma}=\epsilon_{0}=0$, 
$\epsilon_{l}=\epsilon_{r}=0$, $t_{L}=1$, $t_{S}=2$, and $\xi=1$. 
The lead parameters chosen $\epsilon_{\mathcal{L}}=0$ and $t_{\mathcal{L}}=2$.}
\label{trans}
\end{figure*}
%@@@@@@@@@@@@@@@@@@@@@@@@@@@@@@@@@@@@@@@@@@@@@@@@@@@
\begin{equation} 
\begin{aligned}
&{\tilde{\epsilon}}_{\alpha,n}={\tilde{\epsilon}}_{\alpha,n-1}+
\dfrac{\mathcal{X}_{n-1}}{\mathcal{Z}_{n-1}}+
\dfrac{\mathcal{Y}_{n-1}}{\mathcal{Z}_{n-1}}\\
&{\tilde{\epsilon}}_{\beta,n}={\tilde{\epsilon}}_{\alpha,n-1}+
\dfrac{t_{L,n-1}^{2}}{E_{\beta,n-1}}+
\dfrac{\mathcal{Y}_{n-1}}{\mathcal{Z}_{n-1}}\\
&{\tilde{\epsilon}}_{\gamma,n}={\tilde{\epsilon}}_{\gamma,n-1}+
\dfrac{t_{S,n-1}^{2}}{E_{\beta,n-1}}+
\dfrac{\mathcal{X}_{n-1}}{\mathcal{Z}_{n-1}}\\
&{\tilde{\epsilon}}_{l,n}={\tilde{\epsilon}}_{l,n-1}+
\dfrac{\mathcal{X}_{n-1}}{\mathcal{Z}_{n-1}}\\
&{\tilde{\epsilon}}_{r,n}={\tilde{\epsilon}}_{r,n-1}+
\dfrac{\mathcal{Y}_{n-1}}{\mathcal{Z}_{n-1}}\\
&t_{L,n}=\dfrac{t_{L,n-1}^{2}t_{S,n-1}}{\mathcal{Z}_{n-1}},\quad
t_{S,n}=\dfrac{t_{L,n-1}t_{S,n-1}}{E_{\beta,n-1}}
\end{aligned}
\label{recursion2}
\end{equation}
where, $\mathcal{X}_{n-1}=t_{L,n-1}^{2}E_{\gamma,n-1}$, 
$\mathcal{Y}_{n-1}=t_{L,n-1}^{2}E_{\beta,n-1}$, 
$\mathcal{Z}_{n-1}=E_{\beta,n-1}E_{\gamma,n-1}-t_{S,n-1}^{2}$, and 
$E_{j,n-1}=(E-\epsilon_{j,n-1})$ for $j=\alpha,\ \beta,\ \gamma$. 
The transmission coefficient across the effective di-atomic system is 
given by~\cite{stone},
\begin{eqnarray}
&T=\dfrac{4\sin^{2}ka}{\mathcal{A}^{2}+\mathcal{B}^{2}} \nonumber\\
&\text{with,}\quad \mathcal{A}=[(P_{12}-P_{21})+(P_{11}-P_{22})\cos ka] \nonumber\\
&\text{and}\quad \mathcal{B}=[(P_{11}+P_{22})\sin ka]
\end{eqnarray}
where,
$P_{11} =[(E-{\tilde{\epsilon}}_{l}^{\ast})(E-{\tilde{\epsilon}}_{r}^{\ast})]/
t_{L}^{\ast}\tau_{\mathcal{L}}
-t_{L}^{\ast}/\tau_{\mathcal{L}},\ 
P_{12} =-(E-{\tilde{\epsilon}}_{r}^{\ast})/t_{L}^{\ast},\ 
P_{21} =(E-{\tilde{\epsilon}}_{l}^{\ast})/t_{L}^{\ast},\ 
P_{22} =-\tau_{\mathcal{L}}/t_{L}^{\ast}$ are the matrix elements of the transfer 
matrix for the effective di-atomic system, and 
$\cos ka = (E-\epsilon_{\mathcal{L}})/2\tau_{\mathcal{L}}$, `$a$' being 
the lattice constant of the leads and is set to unity throughout the calculation. 

In Fig.~\ref{trans} we have presented the transmission characteristics of a $9$-th 
generation system. 
As we tune the value of $\lambda$ to $\sqrt{3}$, which is the resonance value of 
$\lambda$, the transmittance of the system gets enhanced significantly as apparent 
from Fig.~\ref{trans}(a). This result is at par with the results discussed in 
Sec.~\ref{sec3} and confirms the extended character of the corresponding electronic 
eigenstates. The oscillation in the $T$-$E$ profile is due to the finite dangling chains, 
where a standing wave pattern is formed. In Fig.~\ref{trans}(a) we observe several 
high-transmission zones separated by zero-transmission zones. This feature can be 
useful in designing small scale novel filtering devices with such systems. 
The effect of deviation from the resonance condition on the transmission coefficient 
$T$ is shown in Fig.~\ref{trans}(b), where we have chosen $\lambda=\sqrt{3}/2$, which 
is the $50\%$ deviated value of $\lambda$ from its resonance value. The transmission 
characteristics under this condition exhibit a highly fluctuating behavior indicating 
poor conducting nature of a quasiperiodic system in general. Sudden bursts of 
transmission peaks are due to the formation of isolated extended states as mentioned 
earlier.       
%%%%%%%%%%%%%%%%%%%%%%%%%%%%%%%%%%%%%%%%%%%%%%%%%%%%%%%%%%%%%%%%%%%%%%%%%%%%%%%%%
\section{Conclusion}
\label{sec5}
In conclusion, we have studied the electronic energy spectrum and transmission 
characteristics of a model Fibonacci quasiperiodic system with side-coupled periodic 
QD arrays within a tight-binding framework. The local density of states and 
the transmission coefficient are computed using an RSRG analysis. The major result 
is that, we are able to find a certain relationship between the parameters of the 
Hamiltonian of the system, which triggers absolutely continuous subbands in the 
energy spectrum of the system. The extended nature of the electronic states under 
the suitable resonance condition is affirmed by a prominent enhancement in the transmission 
coefficient. The number and the position of the continuous subbands can be controlled 
by tuning the number of atomic sites in the side-attached chains. This opens up 
a possibility of devising nanoscale electronic filtering devices with such systems. 
We have investigated the effect of side-attachments on a 
Fibonacci type backbone -- this study can be carried over to some other 
quasiperiodic structures like copper mean, Thue-Morse etc. with side-attached 
periodic chains. 

Before we end, it is to be appreciated that the central 
method presented in this communication is applicable to a variety of building 
blocks arranged in a completely disordered fashion. The results indicate a 
unique delocalization of electronic state contrary to the canonical case of 
Anderson localization.         
%%%%%%%%%%%%%%%%%%%%%%%%%%%%%%%%%%%%%%%%%%%%%%%%%%%%%%%%%%%%%%%%%%%%%%%%%%%%%%%%%
\begin{acknowledgements}
The author is thankful to Prof. Arunava Chakrabarti for his illuminating comments 
and suggestions during the preparation of the manuscript and acknowledges DST, 
India for providing funding through an INSPIRE fellowship.
\end{acknowledgements}

\end{document}